\documentclass[11pt,draftclsnofoot, onecolumn,twoside]{IEEEtran}
\usepackage{amsfonts,amssymb,amsbsy,amsmath,paralist,bm,ifthen,color}

\usepackage{graphicx}
\usepackage{subfigure,relsize}

\usepackage[noadjust]{cite}
\usepackage[numbers,sort&compress,square]{natbib}

\usepackage{cases,booktabs}
\usepackage{flushend}

\usepackage{algorithmic,algorithm}

\usepackage{lipsum}

\usepackage{amsthm}

\newtheorem{lem}{Lemma}

\newcommand\Sc{\ensuremath{\mathcal{S}}}

\newcommand\Nc{\ensuremath{{\mathcal{N}}}}

\newcommand\Mc{\ensuremath{{\mathcal{M}}}}
\newcommand\xb{\ensuremath{{\bm x}}}

\newcommand\Ebb{\ensuremath{{\mathbb E}}}
\newcommand\wb{\ensuremath{{\bm w}}}
\newcommand\yb{\ensuremath{{\bm y}}}

\newcommand\sbb{\ensuremath{{\bm s}}}
\renewcommand\sb{\ensuremath{{\bf s}}}
\newcommand\ub{\ensuremath{{\bm u}}}

\newcommand\hb{\ensuremath{{\bm h}}}

\newcommand\Ab{\ensuremath{{\bm A}}}

\newcommand\Cb{\ensuremath{{\bm C}}}

\newcommand\Db{\ensuremath{{\bm D}}}

\newcommand\gb{\ensuremath{{\bm g}}}

\newcommand\pb{\ensuremath{{\bm p}}}

\newcommand\Ub{\ensuremath{{\bm U}}}

\newcommand\rb{\ensuremath{{\bm r}}}

\newcommand\qb{\ensuremath{{\bm q}}}
\newcommand\Vb{\ensuremath{{\bm V}}}

\newcommand\betab{\ensuremath{{\bm \beta}}}

\newcommand\mub{\ensuremath{{\bm \mu}}}

\newcommand\nub{\ensuremath{{\bm \nu}}}

\newcommand\psib{\ensuremath{{\bm \psi}}}

\newcommand\phib{\ensuremath{{\bm \phi}}}

\newcommand\Cbb{\ensuremath{{\mathbb{C}}}}
\newcommand\Rbb{\ensuremath{{\mathbb{R}}}}

\graphicspath{{figures/}}

\begin{document}
	
	\title{Power Minimization in Pinching-Antenna Systems under Probabilistic LoS Blockage}

	\author{Lei~Li, Yanqing~Xu, Tenghao Cai and Tsung-Hui~Chang \vspace{-0.5cm}
		\thanks{\smaller[1] L. Li and T.-H. Chang are with the School of Artificial Intelligence, The Chinese University of Hong Kong, Shenzhen, China, and with the Shenzhen Research Institute of Big Data (email: lei.ap@outlook.com, tsunghui.chang@ieee.org). }
		\thanks{\smaller[1] Y. Xu and T. Cai are with the School of Science and Engineering, The Chinese University of Hong Kong, Shenzhen, China, and with the Shenzhen Research Institute of Big Data (email: xuyanqing@cuhk.edu.cn, 221019048@link.cuhk.edu.cn). 
		}
	}

	\maketitle

	\begin{abstract}
		
		With great flexibility to adjust antenna positions, pinching antennas (PAs) are promising for alleviating large-scale attenuation in wireless networks. In this work, we investigate the antenna positioning and beamforming (AP-BF) design in a multi-PA multi-user system under probabilistic light-of-sight (LoS) blockage and formulate a power minimization problem subject to per-user signal-to-noise ratio (SNR) constraints. For a single PA, we prove the convexity of the simplified problem and obtain its global optimum. For multiple PAs, we derive closed-form BF structures and develop an efficient first-order algorithm to achieve high-quality local solutions. Moreover, the impact of link correlation on the transmit power is analyzed theoretically. Extensive numerical results validate the efficacy of our proposed designs and the substantial performance advantage of PA systems compared with conventional fixed-antenna systems in a term of power saving.
		
		\vspace{0.2cm}
	\end{abstract}

	\IEEEpeerreviewmaketitle

	\vspace{-0.2cm}
	\section{Introduction}
	
	The rapid evolution of wireless communication from 5G toward next-generation (Next-G) networks is driven by the demand for ultra-high data rates, massive connectivity, and pervasive sensing capability \cite{IMT_alliance2024itu}. To meet these ambitious goals, multi-antenna technologies have become one of the critical enablers. In addition, new techniques such as reconfigurable intelligent surfaces (RIS), fluid antennas, and movable antennas further enhance adaptability by reflective phases or adjusting antenna positions to enable more favorable propagation environments. However, they have limited capabilities in mitigating large-scale path loss -- RIS systems suffer from double attenuation, while the limited movement range of fluid antennas and movable antennas within only a few wavelengths restricts their ability to create new line-of-sight (LoS) paths. To overcome these limitations, the emerging concept of pinching antennas (PAs) has drawn increasing attention \cite{tcom25_ding_Flexible, xu2025generalized}. In a PA system, antennas (or pinches) mounted on a dielectric waveguide can be dynamically activated at or moved to arbitrary positions across a broad spatial range, offering enhanced flexibility to align with user locations and establish strong LoS channels.
	
	In PA systems, the antenna positioning (AP) plays a crucial role and has gained great attention in recent studies. The work \cite{tcom25_ding_Flexible} analyzed the ergodic rate of a single-PA time-division multiple access (TDMA) system and showed it surpasses that of the conventional fixed antenna system. Considering a multi-PA single-user system, \cite{wcl25_yx_twostage} proposed a two-stage positioning algorithm to maximize the downlink rate. For a multi-PA multi-user system where each PA over a waveguide is connected to an RF-chain, \cite{wang2025modeling} jointly optimized base station (BS) beamforming (BF) and PA positioning via a penalty-based alternating algorithm to minimize power under signal-to-interference-plus-noise ratio
	(SINR) constraints. Though these works offered valuable insights, they all assumed a deterministic LoS channel between the PA and users, which may not hold in dynamic environments with obstacles and user mobility. To address this, a few works \cite{wcl25_ding_blockage, xu2025pinching, jiang2025spatially} investigated the PA system under probabilistic LoS blockage conditions. {With zero-forcing BF and assuming each PA located closest to its associated user,} the analysis \cite{wcl25_ding_blockage} with revealed that probabilistic LoS blockage largely changes the outage behavior of PA systems. In \cite{xu2025pinching}, PA positioning and beamforming were optimized to maximize the average sum rate of a multi-PA multi-user system via dynamic sampling and alternating optimization. { However, the proposed algorithm is not directly applied to energy-efficient scenarios.}  Additionally, \cite{jiang2025spatially} designed the PA position and power-splitting factor for a PA-assisted simultaneous wireless information and power transfer system {with a single PA and a single user}. While these works mainly focused on communication rate enhancement, power-efficient PA system design under stochastic LoS blockage remains largely unexplored.
	
	To fill the above gap, we study the power minimization under probabilistic LoS blockage in a PA system, where multiple PAs jointly serve multiple users in the TDMA mode. We formulate a joint AP and BF optimization problem that minimizes the transmit power subject to per-user average SNR requirements and PA position constraints. {Unlike existing works that mainly focus on communication-rate maximization (e.g. \cite{xu2025pinching}), the considered design is to satisfy user-specific communication rate requirements with minimum transmit power under stochastic blockage.}  For this problem, we devise efficient first-order algorithms to achieve high-quality solutions. The main contributions are summarized as follows:
	
	{
		\begin{enumerate}[1)]
			\item We first formulate a joint AP-BF optimization problem to minimize the transmit power in a multi-PA multi-user system under probabilistic LoS blockage channels, explicitly capturing stochastic effects overlooked in prior deterministic LoS models. For the single-PA scenario, by exploiting the tightness of constraints, we simplify the original problem into an AP problem and rigorously prove that its convexity, ensuring that the global optimum can be efficiently attained via projected gradient descent (PGD). 
			
			\item For the general multi-PA scenario, we derive optimal beamformers in closed-form as functions of PA positions, achieving a simplified AP problem with far fewer variables. With the derived beamformers, the impact of link correlation is analyzed theoretically.  By leveraging the limited-memory Broyden-Fletcher-Goldfarb-Shanno (L-BFGS) technique \cite{LBFGS_liu1989}, we develop an efficient algorithm for the resulting nonconvex AP problem. Simulation shows that our proposed approach attains significant power savings over conventional fixed-antenna systems. 
			
		\end{enumerate}
	}
	
	\vspace{-0.1cm}
	\section{System Model} \label{sec:model}	
	
	\begin{figure}[t] 
		\centering	
		{\includegraphics[width=0.55\textwidth]{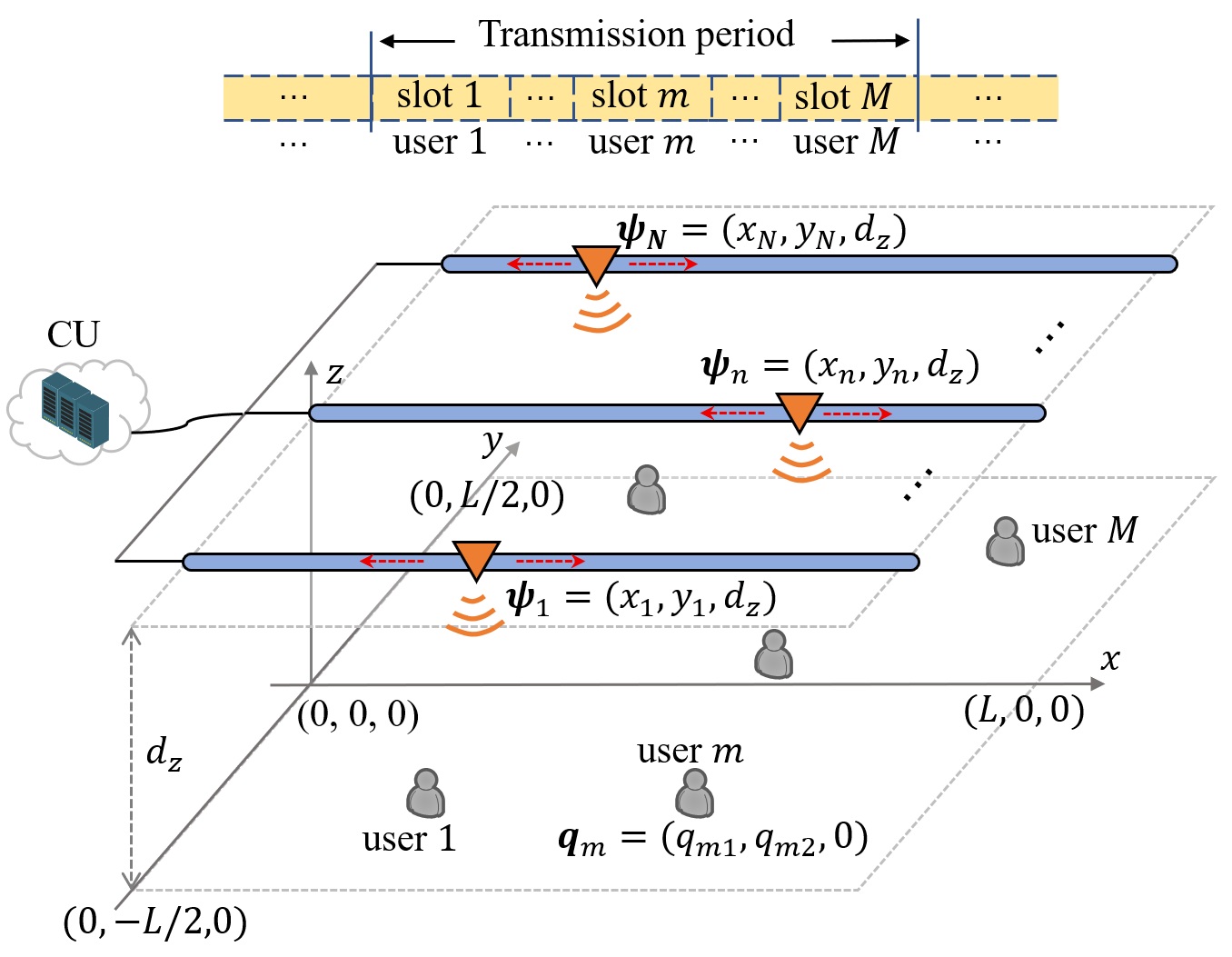}}\vspace{-0.15cm}	
		\caption{A multi-waveguide PA TDMA network with $N$ TPAs, each deployed and movable on an independent waveguide, while all waveguides are connected to a central unit (CU) that coordinates the transmissions to $M$ users.} 
		\label{fig:system} \vspace{-0.35cm}
	\end{figure}
	
	As shown in Fig. \ref{fig:system}, we consider a multi-waveguide PA network, where $N$ transmit PAs (TPAs) deployed on independent waveguides jointly serve $M$ single-antenna communication users. Denote the user set as $\Mc \triangleq \{1, \dots, M\}$ and the TPA set as $\Nc \triangleq \{1, \dots, N\}$. Without loss of generality (w.l.o.g.), the users are assumed to be randomly distributed in a square area of side length $L$ (m). The 3D coordinate of user $m$ is $\qb_m = (q_{m1}, q_{m2}, 0), m \in \Mc$, where $q_{m1} \in [0, L]$ and $q_{m2} \in [\frac{-L}{2}, \frac{L}{2}]$. Each waveguide is deployed parallel to the $x$-axis at height $d_z$, and the waveguides are uniformly spaced above the coverage area. Let $\Delta d_y = L/N$, and denote the feed point of the $n$-th waveguide as $\phib_n = [0, \phi_{y,n}, d_z],  n\in \Nc$, where $\phi_{y,n} = \frac{-L}{2} + (n-\frac{1}{2}) \Delta d_y$. The position of the $n$-th TPA is $\psib_n = [x_{n}, y_n, d_z]^T, n\in \Nc$, where $x_n \in [0, L]$ and $ y_n = \phi_{y,n}$. Each TPA is movable along its associated waveguide with a range much larger than the signal wavelength, enabling flexible placement tailored to user locations.
	
	\vspace{-0.2cm}
	\subsection{Channel Model}
	\vspace{-0.1cm}
	
	According to the classical spherical wave channel model \cite{twc22_pinch_chModel}, the LoS channel between the $n$-th TPA and the $m$-th user is expressed as\footnote{ {Here, the spherical-wave channel model is adopted because the flexible deployment of PAs along potentially long waveguides can yield a large effective aperture and hence a large Rayleigh distance. Meanwhile, PAs are designed to place the radiating point close to users, making the links more likely to be within the Rayleigh distance, for which the spherical-wave model is more appropriate \cite{xu2025generalized}.}}
	\begin{equation} \label{eq:h_los}
		h_{nm}^{\text{L}}(x_n) = {\eta^{1/2} e^{-2 \pi j \big(\frac{\|\psib_n - \qb_m\|}{\lambda} + \frac{\|\psib_n - \phib_n\|}{\lambda_g} \big)}}\big/{\|\psib_n - \qb_m\|},
	\end{equation}
	where $\eta = c^2/(4 \pi f_c)^2$ is a constant determined by the carrier frequency $f_c$ and the speed of light $c$. Here, $\lambda = c/f_c$ denotes the free space wavelength and  $\lambda_g = c/(f_c n_{e}) $ is the waveguide wavelength, where $n_{e}$  represents the effective refractive index of the dielectric waveguide. In \eqref{eq:h_los}, the phase shift includes two components: $e^{-2 \pi j \frac{\|\psib_n - \qb_m\|}{\lambda} } $ from the signal propagation in the free-space and $e^{-2 \pi j + \frac{\|\psib_n - \phib_n\|}{\lambda_g}}$ from the propagation inside the $n$-th waveguide.
	
	In practice, obstacles in the environment like building or furniture, as well as user mobility, introduce randomness into the propagation channel. Accordingly, the existence of the LoS between a TPA and a user is probabilistic. To capature the effect of possible LoS blockage, the existence of the LoS path between TPA $n$ to user $m$ is modeled as a Bernoulli variable $\beta_{nm} \in \{0, 1\}$, where $\beta_{nm} = 1$ and $\beta_{nm} = 0$ indicate the existence and the absence of the LoS link, respectively \cite{wcl25_ding_blockage, xu2025pinching, jiang2025spatially}. For simplicity, we assume that the existence of LoS across different links is independent. { Motivated by analytical blockage studies, the LoS probability from TPA $n$ to user $m$  can be modeled as \cite{tcom25_ding_Flexible, xu2025generalized, twc14_bai_prob} }
	\begin{equation} \label{eq:prob}
		\mathbb{P}(\beta_{nm} = 1) = e^{- \epsilon \|\psib_n - \qb_m\|^2} \triangleq p_{nm},
	\end{equation} where the parameter $\epsilon \in [0.01, 1] \text{m}^{-2}$ \cite{TR_38901} characterizes the obstacle density in the environment. Based on \eqref{eq:h_los} and \eqref{eq:prob}, the instantaneous channel between the $n$-th TPA to user-$m$ is expressed as $h_{nm}(x_n) = \beta_{nm} h_{nm}^{\text{L}}(x_n)$. Further, define $\xb \triangleq [x_1, \dots, x_N] \in \Rbb^N$, {  $\betab_m \triangleq [\beta_{1m}, \dots, \beta_{nm}]^T$ as the vector of LoS indicator variables, and $\hb_m^{\rm L}(\xb) \triangleq [h_{1m}^{\rm L}(x_1), \dots, h_{Nm}^{\rm L}(x_N)]^T \in \Cbb^{N}$.  The composite channel from all TPAs to the $m$-th user can be written as
		\begin{equation}\label{eq:h_miso}
			\hb_{m}(\xb) = \betab_m \odot \hb_m^{\rm L}(\xb) \in \Cbb^{N},
		\end{equation} 
		where $\odot$ denotes the Hadamard product. }
	
	\vspace{-0.3cm}
	\subsection{Communication Model}
	\vspace{-0.1cm}
	
	In this TDMA network, each transmission period is divided into $M$ time slots. W.l.o.g., user $m$ is assumed is served { in the $m$-th slot of length $\tau_m > 0$.} Let $s_m \in \Cbb$ with $\mathbb{E}\{|s_m|^2\}=1$ be the transmit symbol for user $m$, and $\wb_m = [w_{m1}, \dots, w_{mN}]^T \in \Cbb^{N}$ as the corresponding transmit beamformer (TBF) at the TPAs. The received signal at user $m$ is 
	$
	y_m = \hb_{m}^H(\xb) \wb_m s_m + n_m,
	$ 
	where $n_m \sim \mathcal{CN}(0, \sigma_m^2)$ is the additive white Gaussian noise with power $\sigma_m^2$. The instantaneous received SNR at user $m$ is then given by 
	\begin{equation} \label{eq:snr_i}
		\gamma_m = {|\hb_{m}^H(\xb) \wb_m |^2}/{\sigma_m^2}.
	\end{equation}
	Considering the probabilistic LoS blockage, we take the expectation of the received SNR 
	as the performance metric:
	\begin{equation}\label{eq:snr_e}
		\bar \gamma_m (\xb, \wb_m) \triangleq \mathbb{E}_{\{\beta_{nm}\}} \{\gamma_m\}.
	\end{equation}
	
	\vspace{-0.4cm}
	\subsection{Problem Formulation}
	
	In this work, we aim to minimize the total transmit power of the network subject to the per-user expected received SNR requirement, by jointly optimizing the TPA positioning and the TBF. Denote $c_m > 0$ as the minimum expected SNR requirement of user $m$, the AP-BF optimization problem in each transmission period\footnote{Here, the TPA positions are adjusted once per transmission period rather than per slot to avoid frequent antenna movement/activation.} can be formulated as
	\begin{subequations} \label{p:pw_min1}
		\begin{align} \label{eq:obj1}
			\min_{\xb, \{\wb_m\}} &~ \sum_{m \in \Mc} { \tau_m}\|\wb_m\|^2, \\
			\text{s.t.} & ~ \bar \gamma_m (\xb, \wb_m)  \ge c_m, \forall m \in \Mc, \label{eq:cons_snr}
			\\
			&~ 0 \le x_n \le L, \forall n \in \Nc. \label{eq:cons_TPA}
		\end{align}
	\end{subequations}
	While the objective \eqref{eq:obj1} and the position constraints \eqref{eq:cons_TPA}  of problem \eqref{p:pw_min1} are convex, the TPA positioning variables $\xb$ are strongly coupled in the SNR constraints \eqref{eq:cons_snr}, as seen from \eqref{eq:snr_i}. Additionally,  one can observe from \eqref{eq:h_los} that the TPA positions affect both the signal phase in the exponential term as well as the path loss in the denominator of $h_{nm}^{\text{L}}(x_n)$, making $\gamma_m $ a highly non-convex function. Moreover, the expectation operation $\mathbb{E}_{\beta_{nm}}\{\cdot\}$ in \eqref{eq:snr_e} further complicates the expression of $\bar \gamma_m (\xb, \wb_m)$. These factors collectively make \eqref{p:pw_min1} challenging to solve. In the following, we first consider the joint AP-BF optimization in the single-PA scenario and develop an efficient algorithm. Afterward, we will investigate the design in the general multi-PA scenario.   
	
	\vspace{-0.3cm}
	\section{Global optimal solution in single-PA scenario} \label{sec:single}
	
	In this section, we focus on a special scenario where only a single TPA is deployed, i.e., $N=1$. In this scenario, the BF optimization in problem \eqref{p:pw_min1} simplifies to power allocation. To handle it,  we first derive the expressions of the expected SNR $\{\bar \gamma_m(\cdot)\}$, based on which, an efficient joint AP and power allocation algorithm is devised to find the optimal solution of the simplified problem.   
	
	For brevity, we omit the TPA index $n$ since $N=1$. Based on \eqref{eq:h_los} and \eqref{eq:h_miso}, the channel from the TPA to the $m$-th user will be $
	h_m(x) = \beta_{nm} h_m^{\rm L}(x) ,
	$
	where $h_m^{\rm L}(x) = \eta^{1/2} e^{-2 \pi j \big(\frac{\|\psib - \qb_m\|}{\lambda} + \frac{\|\psib - \phib\|}{\lambda_g} \big)}/\|\psib - \qb_m\| $ with $\psib = [x, 0, d_z]^T$. Accordingly, the instantaneous received SNR at user $m$ is $\gamma_m = \frac{P_m|h_m(x)|^2}{\sigma_m^2},$
	where $P_m = |w_m|^2$ is the associated transmit power. Then, the expectation of $\gamma_m$ is derived as
	\begin{equation} \label{eq:E_snr_s}
		\begin{aligned}
			\mathbb{E}_{\beta_{m}} \{\gamma_m\} 
			& = P_m/\sigma_m^2 \mathbb{E}_{\beta_{nm}} \{ \beta_{nm} \}  |h_m^{\rm L}(x)|^2, \\
			& = \frac{P_m \eta}{\sigma_m^2} \frac{e^{- \epsilon \|\psib - \qb_m\|^2}}{\|\psib - \qb_m\|^2}, \\
			& = \frac{P_m \eta}{\sigma_m^2} \frac{e^{- \epsilon [(x-q_{m1})^2 + z_m]}}{(x-q_{m1})^2 + z_m} \triangleq \bar \gamma_m(x, P_m ),
		\end{aligned}
	\end{equation}
	where $z_m = q_{m2}^2 + d_z^2$ is a constant. Based on \eqref{eq:E_snr_s}, problem \eqref{p:pw_min1} under $N=1$ can be simplified to 
	\begin{subequations} \label{p2:pw_min}
		\begin{align} \label{eq:obj2}
			\min_{x, \{P_m\}} &~\sum_{m \in \Mc}  { \tau_m} P_m, \\
			\text{s.t.} & ~ \bar\gamma_m(x, P_m )  \ge c_m, \forall m \in \Mc, \label{eq:cons_snr2}
			\\
			&~ 0 \le x \le L. \label{eq:cons_TPA2}
		\end{align}
	\end{subequations}
	For problem \eqref{p2:pw_min}, it is straightforward to verify that its optimal solution will satisfy \eqref{eq:cons_snr2} with equality, i.e., $\bar\gamma_m(x, P_m )  = c_m, \forall m \in \Mc$. 
	Based on this property, for a fixed $x$, the optimal transmit power $\{P_m\}$ admits a closed-form expression as
	\begin{equation} \label{eq:pw}
		P_m(x) = \tilde{c}_m [(x-q_{m1})^2 + z_m]  e^{\epsilon [(x-q_{m1})^2 + z_m]}, \forall m \in \Mc,\!\!
	\end{equation}
	where $\tilde{c}_m = \sigma_m^2 c_m/\eta$.
	This result greatly reduces the number of variables and constraints of problem \eqref{p2:pw_min}. Substituting \eqref{eq:pw} into \eqref{p2:pw_min}, this problem is further simplified to 
	\begin{equation} \label{p3:pw_min}
		\begin{aligned}
			\min_{0 \le x \le L} &~ \sum\nolimits_{m \in \Mc}  { \tau_m} P_m(x).
		\end{aligned}
	\end{equation}
	So far, problem \eqref{p3:pw_min} has one remaining variable $x$ with linear constraints, while the difficulty to solve it lies in the intricate $P_m(x)$ in its objective function. To handle it, by examining the structure of $P_m(x)$, we establish the following Lemma \ref{Lemma:1}.
	\begin{lem} \label{Lemma:1} 
		The function $P_m(x)$ is convex w.r.t. $x$.
	\end{lem} 
	\begin{proof} 
		For brevity, define $t_m(x) \triangleq (x-q_{m1})^2 + z_m$, such that $P_m(x) = \psi(t_m(x))$ with $ \psi(t_m) = \tilde{c}_m t_m e^{\epsilon t_m}$. Let us first examine the convexity of $\psi(t_m)$. Specifically, the first-order derivative of  $\psi(t)$ is given by
		$
		\psi'(t_m) =  \tilde{c}_m e^{\epsilon t_m} (1+ \epsilon t_m),
		$
		and the second-order derivative can be derived as
		$
		\psi''(t_m) = \tilde{c}_m e^{\epsilon t_m} \epsilon(2+ \epsilon t_m).
		$
		Since $t_m(x) > 0$, $\psi'(t_m) > 0 $ and $\psi''(t_m) > 0 $ will hold. Therefore, $\psi(t)$ is a strongly convex function for $t > 0$. Moreover, $t_m(x)$ is a convex function w.r.t. $x$. According to the composition rule \cite{boyd2004convex}, $P_m(x)$ will be convex w.r.t. $x$.
	\end{proof}

	Based on Lemma \ref{Lemma:1}, problem \eqref{p3:pw_min} can be efficiently solved by PGD. Specifically, in each iteration $\ell$, $x$ is updated by $x^{\ell +1} \leftarrow \mathsf{P}_x^\ell \big[x^\ell - \bar s_1\sum\nolimits_{m \in \Mc} \nabla_x P_m(x^\ell) \big],$
	where $\bar s_1 > 0$ is the step size and $\mathsf{P}_x[\cdot]$ is the projection operator over $[0, L]$. Moreover, PGD will converge to a global optimum $x^\star$ of this problem. Substituting $x^\star$ into \eqref{eq:pw} then yields the optimal transmit power for each user as $P_m^\star = P_m(x^\star), m \in \Mc$.

	\section{Joint AP-BF design in multi-PA scenario}
	\label{sec:alg_multi}
	
	In the general scenario of $N>1$ TPAs, the AP-BF optimization problem \eqref{p:pw_min1} gets significantly more challenging, as $\bar \gamma_m (\xb, \wb_m) $ in \eqref{eq:cons_snr} involve more variables that are strongly coupled. To handle it, we first derive its explicit expression.

	Define $\Ab_m(\xb) \triangleq \mathbb{E}_{\{\beta_{nm}\}}[\hb_{m}(\xb)\hb_{m}^H(\xb)]$ as the channel covariance matrix. 
	From \eqref{eq:snr_i} and \eqref{eq:snr_e}, 
	$\bar \gamma_m (\xb, \wb_m) = \wb_m^H \Ab_m(\xb) \wb_m/\sigma_m^2$. Moreover, based on \eqref{eq:h_miso}, 
	\begin{equation} \label{eq:C_h}
		\begin{aligned}
			\Ab_m(\xb) 
			& = \mathbb{E}_{\{\beta_{nm}\}}[(\betab_m \betab_m^T)] \odot [\hb_m^{\rm L}(\xb) {(\hb_m^{\rm L}(\xb))}^H], \\
			& \triangleq \Cb_{\betab_m}  \odot \Cb_{h^L},
		\end{aligned}
	\end{equation}
	where $\Cb_{\betab_m} \triangleq \mathbb{E}_{\{\beta_{nm}\}}[(\betab_m \betab_m^T)] $ is the covariance matrix of the LoS indicator variables.
	With the assumption of independent LoS blockage across links, its $(i,j)$-th entry is given by
	\begin{equation} \label{eq:C_beta}
		[\Cb_{\betab_m}]_{ij} = \Ebb\{ \beta_{im}\beta_{jm} \} = \begin{cases}
			p_{im}, &  i  = j, \\
			p_{im} p_{jm}, & i \neq j.
		\end{cases}
	\end{equation}
	Define $\pb_m \triangleq [p_{1m}(x_1),\dots, p_{Nm}(x_N)]$. Substituting \eqref{eq:C_beta} into \eqref{eq:C_h} yields
	\begin{subequations} \label{eq:E_S2}
		\begin{align}
			\Ab_m(\xb) 
			&= [\pb_m\pb_m^T + \text{diag}(\pb_m \odot (\mathbf{1} -\pb_m))]  \\
			& ~~~\odot [\hb_m^{\rm L}(\xb) {(\hb_m^{\rm L}(\xb))}^H ], \\
			& = \ub_m \ub_m^H + \Vb_m \succ \mathbf{0}, \label{eq:E_S2_2_T1}
		\end{align}
	\end{subequations}
	where $\ub_m \triangleq [p_{1m} h_{1m}^{\rm L}, \dots, p_{Nm} h_{Nm}^{\rm L}]^T \in \Cbb^{N}$ and  $\Vb_m \triangleq \text{diag}(\{p_{im}(1-p_{im}) |h_{im}^{\rm L}|^2\}_{i=1}^N)$.
	In \eqref{eq:E_S2_2_T1}, the first term represents the coherent combination gain of signals from different TPAs, while the second term captures the power variation due to LoS link uncertainty as a weighted non-coherent sum.  This decomposition reveals the impact of each component on the received power clearly. 
	
	With the above results, 
	the AP-BF optimization problem \eqref{p:pw_min1} is updated to
	\begin{subequations} \label{p:AP_BF1}
		\begin{align} \label{eq:obj_APBF1}
			\min_{\xb, \{\wb_m\}} &~ \sum_{m \in \Mc}  { \tau_m} \|\wb_m\|^2, \\
			\text{s.t.} & ~ \frac{ \wb_m^H \Ab_m(\xb) \wb_m}{\sigma_m^2}  \ge c_m, \forall m \in \Mc, \label{eq:cons_snr_APBF1}
			\\
			&~ 0 \le x_n \le L, \forall n \in \Nc. \label{eq:cons_TPA_APBF1}
		\end{align}
	\end{subequations}
	Given $\xb$, the BF optimization in \eqref{p:AP_BF1} can be decoupled into $M$ subproblems, and the $m$-th subproblem is equivalent to
	\begin{subequations} \label{p:cons_snr_BF}
		\begin{align}
			\min_{\wb_m \in \Cbb^N} &~ \|\wb_m\|^2, \\
			\text{s.t.} & ~ \wb_m^H \Ab_m(\xb) \wb_m  \ge c_m {\sigma_m^2}. \label{eq:cons_snr_BF}
		\end{align}
	\end{subequations}
	Introduce an intermediate variable $\mub_m \in \Cbb^N$ satisfying $\|\mub_m\| = 1$ and $\wb_m = \alpha_m \mub_m$, where $\alpha_m > 0$ is a scaling factor. Then, the smallest $\alpha_m$ meeting \eqref{eq:cons_snr_BF} should satisfy $\alpha_m^2 = \frac{c_m {\sigma_m^2}}{\mub_m^H \Ab_m(\xb) \mub_m }$. Therefore, problem \eqref{p:cons_snr_BF} is equivalent to 
	\begin{equation} \label{p:BF_eig}
		\min_{ \|\mub_m\|=1} ~ \frac{c_m {\sigma_m^2}}{\mub_m^H \Ab_m(\xb) \mub_m }.
	\end{equation}
	Obviously, this optimal solution of \eqref{p:BF_eig} is the principal eigenvector $\nub_m (\Ab_m(\xb))$ of $\Ab_m(\xb)$. Denote the largest eigenvalue of $\Ab_m(\xb)$ as $\lambda_m^{\rm max} (\Ab_m(\xb))$. The optimal $\mub_m^* = \nub_m (\Ab_m(\xb))$ and $\alpha_m^* = \sqrt{c_m {\sigma_m^2}/\lambda_m^{\rm max} (\Ab_m(\xb))}$. Therefore, the optimal solution of \eqref{p:cons_snr_BF} is given by
	\begin{equation} \label{eq:w_m_opt}
		\wb_m^\star(\xb) = \sqrt{\frac{c_m \sigma_m^2}{\lambda_m^{\rm max} (\Ab_m(\xb))} } \nub_m (\Ab_m(\xb)).
	\end{equation}
	
	Substituting \eqref{eq:w_m_opt} into \eqref{p:AP_BF1}, the AP-BF problem \eqref{p:AP_BF1} will be reduced to the following problem 
	\begin{subequations} \label{p:AP_BF2}
		\begin{align} \label{eq:obj_APBF2}
			\min_{\xb} &~  \sum_{m \in \Mc}  { \tau_m} \frac{c_m \sigma_m^2}{\lambda_m^{\rm max} (\Ab_m(\xb))} \triangleq f(\xb) \\
			\text{s.t.}
			&~ 0 \le x_n \le L, \forall n \in \Nc. \label{eq:cons_TPA_APBF2}
		\end{align}
	\end{subequations}
	{ Notably, $f(\xb)$ is impacted by the channel covariance matrix $\Ab_m(\xb)$, which has different values under link correlation. Specifically, denote $\Ab_m'(\xb)$ and $\Cb_{\betab_m}'$ as the channel covariance matrix and the covariance matrix of the LoS indicator variables $\betab_m$ in the correlated case, respectively. Similar to \eqref{eq:C_h}, $\Ab_m'(\xb) = \Cb_{\betab_m}'  \odot \Cb_{h^L}$, but $\Cb_{\betab_m}'$ change to
		\begin{equation} \label{eq:C_beta_corr}
			[\Cb_{\betab_m}']_{ij} = \Ebb\{ \beta_{im}\beta_{jm} \} = \begin{cases}
				p_{im}, &  i  = j, \\
				p_{im}p_{jm} + \rho_{ij} \delta_{im} \delta_{jm}, & i \neq j,
			\end{cases}
		\end{equation}
		where $\rho_{ij}\in[-1, 1]$ is the correlation coefficient between $\beta_{im}$ and $\beta_{jm}$, and $\delta_{im} = \sqrt{p_{im} (1- p_{im})}$ is the standard deviation of Bernoulli r.v. $\beta_{im}$. Then, the following lemma holds. }
	{
		\begin{lem} \label{Lemma:2}
			If all correlation coefficients $\rho_{ij} \ge 0$ and at least one $\rho_{ij} >0$, then $\lambda_{m}^{\max}(\Ab_m'(\xb)) \ge \lambda_{m}^{\max}(\Ab_m(\xb))$. Therefore, with the beamformer $\wb_m^\star(\xb)$ in \eqref{eq:w_m_opt}, the power consumption is lower in the positively correlated scenario.     
		\end{lem}
		\begin{proof}
			The details of the proof are given in Appendix \ref{app:1}.
		\end{proof}
		
	}
	
	Compared to \eqref{p:AP_BF1}, problem \eqref{p:AP_BF2} only needs to optimize the AP with the linear inequality constraints. However, the objective \eqref{eq:obj_APBF2} remains highly intricate due to the denominator of each term. While $\lambda_m^{\rm max} (\Ab_m(\xb))$ is convex w.r.t. $\Ab_m(\xb)$, its reciprocal is neither convex nor concave. Moreover, the components of $\Ab_m(\xb)$ -- namely, $\ub_m$ and $\Vb_m$ -- are highly non-linear mapping from $\xb$, as evident from \eqref{eq:E_S2}. Consequently, $f(\xb)$ is a highly non-convex function. Additionally, due to the complex structure of $\Ab_m(\xb)$, it is difficult to construct a tractable convex upper bound for $f(\xb)$ and obtain the global optimum of \eqref{p:AP_BF2}. 
	
	In this work, we instead seek to find a high-quality local solution by exploiting the limited-memory L-BFGS \cite{LBFGS_liu1989} technique, which is a limited-memory variant of the quasi-Newton method. Compared with gradient descent, L-BFGS  converges faster, and by iteratively approximating the inverse Hessian only with a small history of updates, it greatly reduces the computational and storage burden relative to full Newton methods. Specifically, set a memory buffer of size $I_b$. At iteration $k$, a search direction $\pb^k$ is computed based on the gradient $\nabla_{\xb}f(\xb)$ and the intermediate parameters $\{(\bar \sbb_i, \bar \yb_i, \rho_i)\}_{i= 1}^{I_b}$ stored in the buffer that approximate the inverse Hessian. A line search is then conducted along that direction to determine the step size $t_k$ to update the solution, followed by updating the buffer intermediate parameters. Following this procedure, the proposed AP-BF algorithm based on projected L-BFGS is detailed in Algorithm \ref{alg:AP_BF}.
	
	{ {\textit{Complexity Analysis}: }   Algorithm \ref{alg:AP_BF} takes only first-order computations, with the main cost being evaluating $\nabla_{\xb} f(\xb) = -\sum_{m=1}^M \frac{ \tau_m c_m \sigma_m^2}{(\lambda_m^{\rm max})^2 } \nub_m^H \nabla_{\xb} \Ab_m(\xb) \nub_m$, based on which the per-iteration complexity of our proposed algorithm can be derived as $\mathcal{O}\big(MN(41 + 10K_{\text{eig}})\big)$, where $K_{\text{eig}}$ denotes  the number of iterations taken by the power method to compute $\nub_m$. Hence, Algorithm \ref{alg:AP_BF} has polynomial complexity and is efficient.}

	\begin{figure}[t]
		\vspace{-0.25cm}
		\begin{algorithm}[H]
			\begin{small}
				\caption{Proposed AP-BF Algorithm}
				\begin{algorithmic}[1] \label{alg:AP_BF}
					\STATE Initialize TPA locations $\{x_n^0\}_{n=1}^N$, $I_b$, intermediate parameters $\Sc \triangleq \{(\bar \sbb_i, \bar\yb_i, \rho_i)\}_{i= 1}^{I_b}$, and the iteration index $k \leftarrow 0$.
					\REPEAT
					\STATE Compute the gradient $\gb^k = \nabla_{\xb} f(\xb^k)$ and initialize $\bar\qb \leftarrow \gb^k$. 
					\REPEAT
					\STATE Update $\tau_i \leftarrow \rho_i \bar \sbb_i^T \bar\qb$, $ \bar\qb \leftarrow \bar\qb - \tau_i \bar\yb_i$.
					\STATE $i \leftarrow i-1$.
					\UNTIL $i < 1$.
					\STATE Update $\gamma_k = \bar \sbb_{k-1}^T \bar\yb_{k-1}/\bar\yb_{k-1}^T \bar\yb_{k-1}$, $\rb \leftarrow \gamma_k \bar\qb$.
					\STATE $i \leftarrow i+1$.
					\REPEAT
					\STATE Update $\zeta_i \leftarrow \rho_i \bar\yb_i^T \rb, \rb \leftarrow \rb + \bar \sbb_i (\tau_i - \zeta_i)$.
					\STATE $i \leftarrow i+1$.
					\UNTIL $i > I_b$.
					\STATE Compute the search direction by $\pb^k = -\rb$.
					\STATE Conduct backtracking line search to find a proper step size $t_k$.
					\STATE Update $\xb^{k+1} \leftarrow \mathsf{P}_x [\xb^k + t_k \pb^k]$.
					\STATE Compute $\bar \sbb_k = \xb^{k+1} - \xb^k, \bar\yb_k = \gb^{k+1} - \gb^k$ and $\rho_k = 1/(\bar\yb_k^T \sb_k)$.
					\STATE Update $\Sc$ by $(\bar \sbb_i, \bar\yb_i, \rho_i) \leftarrow (\bar \sbb_{i+1}, \bar\yb_{i+1}, \rho_{i+1}), i = 1, \dots, I_b -1$ and $ (\bar \sbb_{I_b}, \bar\yb_{I_b}, \rho_{I_b}) \leftarrow (\bar \sbb_k, \bar\yb_k, \rho_k)$. 
					\STATE Update $k \leftarrow k+1$.
					\UNTIL a predefined stopping criterion is satisfied.
					\STATE {\bf Output} $\{ \xb^k\}$, and compute $\wb_m^\star(\xb^k)$ by \eqref{eq:w_m_opt}.
				\end{algorithmic}\vspace{-0.1cm}
			\end{small}
		\end{algorithm} 
		\vspace{-0.75cm}
	\end{figure}

	\vspace{-0.0cm}
	\section{Simulation Results \label{sec:simu}}
	\vspace{-0.0cm}
	
	In this section, our proposed designs are evaluated by numerical simulations. In the simulation, $M$ users are randomly sampled in a $L \times L$ area with $L = 20~\text{m}$. The height of waveguides is set to $d_z = 10~\text{m}$. The carrier frequency $f_c = 28~\text{GHz}$, the effective refractive index $n_e = 1.4$, the noise power $\sigma_m^2=10^{-14}~\text{watt}$ and $\tau_m=1$. { For comparison, we include two benchmark schemes with fixed PA positions. In the `Center' scheme, the $n$-th PA is activated at the center of waveguide-$n$, which also serves as the initial point for our proposed algorithm. In the `Random' scheme, PAs are randomly activated along the waveguides. For both benchmarks, given the PA positions, beamformers are optimized via \eqref{eq:w_m_opt}. }

	\begin{figure}[t]
		\centering
		{\resizebox{0.55\textwidth}{!}
			{\includegraphics{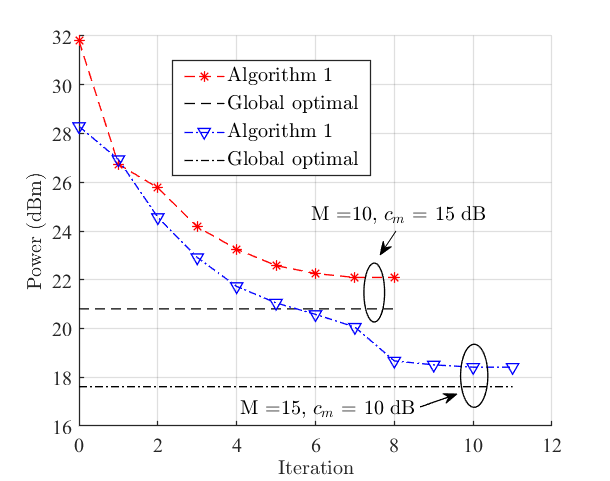}}}	\vspace{-0.3cm} 
		\caption{Convergence behavior under different network configurations, $N=4$ TPAs and $\epsilon = 0.05$.} 
		\label{fig:cvg}\vspace{-0.3cm}
	\end{figure}
	
	\begin{figure}[t]
		\centering
		{\resizebox{0.55\textwidth}{!}
			{\includegraphics{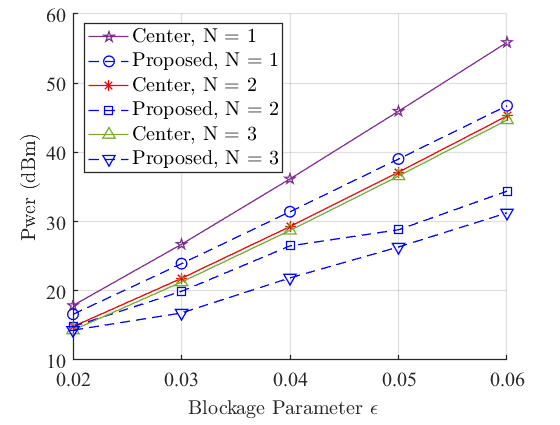}}}	\vspace{-0.3cm} 
		\caption{Power versus blockage parameter $\epsilon$ under different number of TPAs. $M=10$ and $c_m = 20$ dB.} 
		\label{fig:pw_beta}\vspace{-0.3cm}
	\end{figure}

	\begin{figure}[t]
		\centering
		{\resizebox{0.55\textwidth}{!}
			{\includegraphics{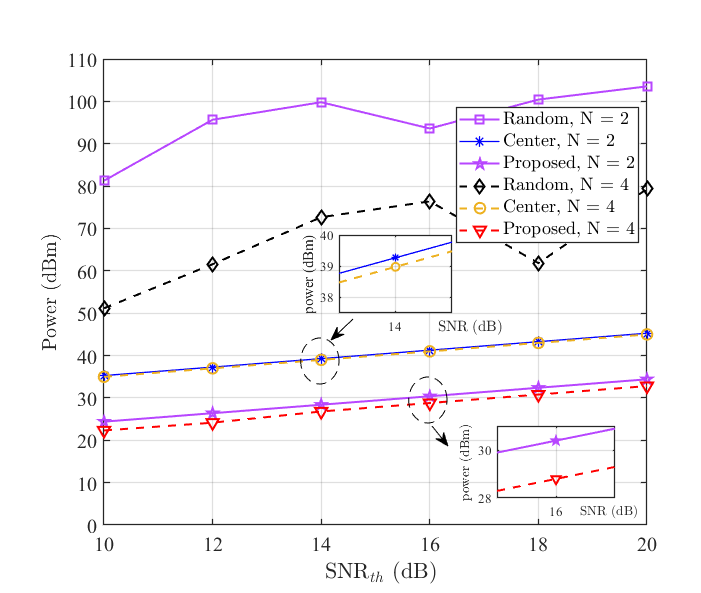}}}	\vspace{-0.3cm} 
		\caption{ Power versus SNR threshold $c_m$ under different number of TPAs. $M=10$ and $\epsilon = 0.06$.} 
		\label{fig:pw_snr}\vspace{-0.3cm}
	\end{figure}

	\begin{figure}[htbp] 
		\centering	
		\subfigure[Users near the left side.]		
		{\includegraphics[width=0.48\textwidth]{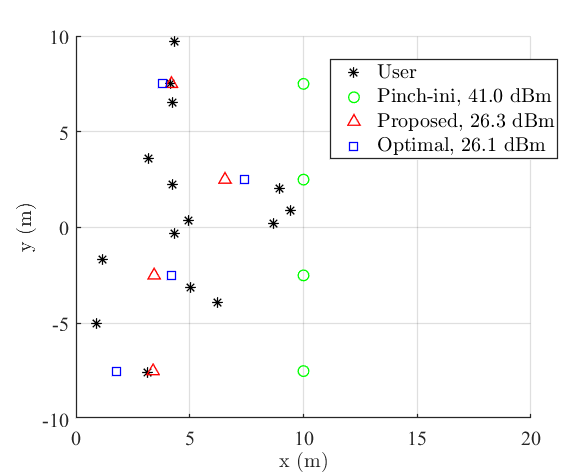}}
		\subfigure[Users near the diagonal.]	
		{\includegraphics[width=0.48\textwidth]{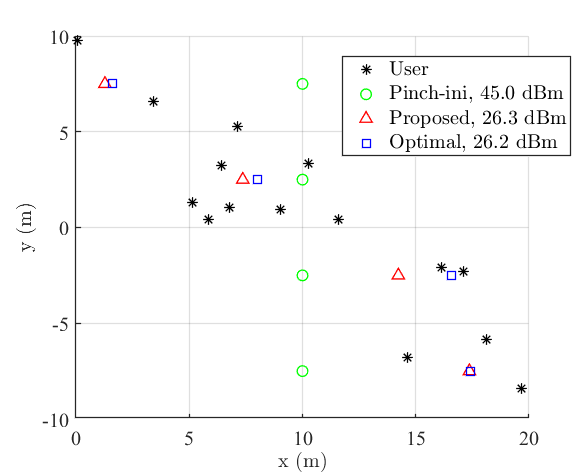}}			
		\caption{The optimized TPA locations under different user spatial distributions. $\epsilon = 0.06$ and $c_m = 15~\text{dB}$} 
		\label{fig:pw_UE}
		\vspace{-0.3cm}
	\end{figure}

	Fig. \ref{fig:cvg} showcases the convergence behavior of our proposed Algorithm \ref{alg:AP_BF}, with the global optimal power obtained by exhaustive antenna position search. It can be observed that Algorithm \ref{alg:AP_BF} converges to a stationary point within only a few iterations under different configurations of the number of users $M$ and SNR thresholds $c_m$. Meanwhile, the power consumption is effectively reduced and the optimized value approaches the global optimum.

	{ Fig. \ref{fig:pw_beta} demonstrates the achieved power versus the LoS blockage parameter $\epsilon$.  The power of all algorithms increases with a larger $\epsilon$, which is expected since a larger $\epsilon$ indicates denser obstacles (cf. \eqref{eq:prob}), making LoS channels from TPAs to users more prone to blockage. For the same $\epsilon$ and $N$, our proposed scheme achieves substantial power reduction over the `Center' benchmark, with greater savings under denser obstacles. This is because the reconfigurable TPA positions can adapt to blockage conditions and user distributions, thereby establishing more favorable channel paths with lower attenuation. Notably, our proposed scheme consistently delivers distinct power savings under different blockage levels.}
	
	{ In Fig. \ref{fig:pw_snr} evaluates the impact of the SNR threshold $c_m$ under different numbers of TPAs $N$. As expected, higher SNR requirements lead to more power consumption for both our proposed algorithm and the `Center' scheme. Nevertheless, our design consistently achieves lower power consumption than the benchmarks, particularly outperforming the `Random' scheme, where randomly placed TPAs may be located far from users and thus severely degrade performance. }  
	
	{ 
		In Fig. \ref{fig:pw_UE}, the TPA locations obtained by our proposed scheme alongside the optimal locations via exhaustive search are shown under different user spatial distributions with $M=15$. Whether users are distributed near the left side or along the diagonal, the TPA positions from our algorithm align well with the user distribution and closely match the optimal ones. This behavior indicates that TPAs should be activated near users to maintain short distances, thereby increasing LoS probability. The results validate that our proposed design effectively adapts TPA positions to the network topology, achieving significant power savings in TPA systems.
	}

	\vspace{-0.0cm}
	\section{Conclusion \label{sec:conclusion}}
	
	In this work, we investigated the AP-BF optimization in a multi-PA multi-user network under stochastic blockage conditions from the power saving perspective. For the single-PA scenario, we have proved the convexity of the problem and achieved its global optimum by the first-order PGD. For the general PA scenario, we derived closed-form solutions of the beamformers, which greatly reduce the number of variables, leading to a simplified multi-PA positioning problem. { For the resultant problem, we have analyzed the impact of link correlation and developed an efficient L-BFGS-based algorithm that only takes first-order computations.} Our proposed algorithms can achieve a distinct power reduction, offering a promising solution for the energy-efficient designs in PA systems. 
	
	\begin{appendices}
		\section{Proof of Lemma \ref{Lemma:2}} \label{app:1}
		First, $h_{im}^{\rm L}$ can be expressed as $h_{im}^{\rm L} = |h_{im}^{\rm L}|e^{j \theta_{im}}$ with $\theta_{im} = \angle h_{im}^{\rm L}, i \in \Nc$. Define $\Db_h \triangleq \text{diag}(h_{1m}^{\rm L},\dots, h_{Nm}^{\rm L})$. $\Db_{|h|} \triangleq \text{diag}(|h_{1m}^{\rm L}|,\dots, |h_{Nm}^{\rm L}|)$ and $\Ub \triangleq \text{diag}(e^{j\theta_{1m}}, \dots, e^{j\theta_{Nm}})$. Then, we have $\Ab_m(\xb)  = \Db_h \Cb_{\betab_m} \Db_h^H$ and $\Ab_m'(\xb) = \Db_h \Cb_{\betab_m}' \Db_h^H $ .
		
		Notice that
		\begin{equation}
			\begin{aligned}
				\Ab_m(\xb) = \Db_{|h|}\Ub \Cb_{\betab_m} \Ub^H \Db_{|h|}.
			\end{aligned}
		\end{equation}
		By the definition of Rayleigh quotient, it can be verified that $\lambda_{m}^{\max}(\Ab_m(\xb)) = \lambda_{m}^{\max}(\Db_{|h|} \Cb_{\betab_m}(\xb) \Db_{|h|})$. Similarly, $\lambda_{m}^{\max}(\Ab_m'(\xb)) = \lambda_{m}^{\max}(\Db_{|h|} \Cb_{\betab_m}'(\xb) \Db_{|h|})$. Therefore, it suffices to compare the largest eigenvalues of $\Db_{|h|} \Cb_{\betab_m}(\xb) \Db_{|h|}$ and $\Db_{|h|} \Cb_{\betab_m}'(\xb) \Db_{|h|}$, both of which have all non-negative entries. 
		
		From the expressions of $\Cb_{\betab_m}(\xb)$ in \eqref{eq:C_beta} and $\Cb_{\betab_m}'(\xb)$ in \eqref{eq:C_beta_corr},  it follows that
		\begin{equation} \label{eq:matrix_greater}
			\Db_{|h|} \Cb_{\betab_m}'(\xb) \Db_{|h|} \succeq \Db_{|h|} \Cb_{\betab_m}(\xb) \Db_{|h|}, 
		\end{equation}
		where `$\succeq$' denotes `elementwise greater than or equal to'. According to Perron-Frobenius Theorem, for two non-negative matrices that satisfy \eqref{eq:matrix_greater}, their largest eigenvalues satisfy
		\begin{equation}
			\lambda_{m}^{\max}(\Db_{|h|} \Cb_{\betab_m}'(\xb) \Db_{|h|}) \ge \lambda_{m}^{\max}(\Db_{|h|} \Cb_{\betab_m}(\xb) \Db_{|h|}).
		\end{equation}
		Consequently, $\lambda_{m}^{\max}(\Ab_m'(\xb)) \ge \lambda_{m}^{\max}(\Ab_m(\xb)).$ Substituting this result into $f(\xb)$ in \eqref{eq:obj_APBF2} yields a smaller objective value in the scenario of positively correlated links.
		
	\end{appendices}
	
	\vspace{0.2cm}
	\footnotesize 
	\bibliographystyle{IEEEtran}
	\bibliography{refs_pinchPw2}
	
	\ifCLASSOPTIONcaptionsoff
	\newpage
	\fi

\end{document}